# Applying Machine Learning to Understand Water Security and Water Access Inequality in Underserved Colonia Communities

Zhining Gu[1], Wenwen Li[1,*], Michael Hanemann[2], Yushiou Tsai[3], Amber Wutich[3], Paul Westerhoff[4], Laura Landes[5], Anais D. Roque[3], Madeleine Zheng[6], Carmen A. Velasco[4], Sarah Porter[7]

[1] School of Geographical Sciences and Urban Planning, Arizona State University
[2] School of Sustainability, Arizona State University
[3] School of Human Evolution and Social Change, Arizona State University
[4] School of Sustainable Engineering and the Built Environment, Arizona State University
[5] Rural Community Assistance Partnership
[6] Biodesign Center for Fundamental & Applied Microbiomics, Arizona State University
[7] The Kyl Center for Water Policy, Arizona State University

*Corresponding Author: wenwen@asu.edu

**Abstract**

This paper explores the application of machine learning to enhance our understanding of water accessibility issues in underserved communities called *colonias* located along the northern part of the United States–Mexico border. We analyzed more than 2000 such communities using data from the Rural Community Assistance Partnership (RCAP) and applied hierarchical clustering and the adaptive affinity propagation algorithm to automatically group colonias into clusters with different water access conditions. The Gower distance was introduced to make the algorithm capable of processing complex datasets containing both categorical and numerical attributes. To better understand and explain the clustering results derived from the machine learning process, we further applied a decision tree analysis algorithm to associate the input data with the derived clusters, to identify and rank the importance of factors that characterize different water access conditions in each cluster. Our results complement experts' priority rankings of water infrastructure needs, providing a more in-depth view of the water insecurity challenges that the colonias suffer from. As an automated and reproducible workflow combining a series of tools, the proposed machine learning pipeline represents an operationalized solution for conducting data-driven analysis to understand water access inequality. This pipeline can be adapted to analyze different datasets and decision scenarios.

**Keywords:** machine learning, water security, hierarchical clustering, adaptive affinity propagation, underserved communities

## 1. Introduction

Colonias, which are located along the United States (US)–Mexico border, often suffer from water insecurity, demonstrating a stark contrast with large US cities that have sophisticated drinking and wastewater systems (Wescoat, 2007). Safe water is very often unavailable in colonias due to their geographical isolation, neglectful water governance, and the lack of infrastructure resources. The colonia communities, which have small populations, often lack a sufficient user base to finance the capital and operational costs of typical centralized delivery systems for water supply and wastewater treatment. To compensate for the lack of sewage systems, the vast majority of colonia residents have had to build their own septic tank systems, which often fail to comply with standard health and construction codes (Rios & Meyer, 2009). Moreover, of the meager amount of water that is accessible to the colonia residents, much is contaminated with chemicals (Rios & Meyer, 2009; Rios-Arana et al., 2004; Balazs et al., 2011; Moore et al., 2011) or microorganisms (Margarita Moya, 2017; Travis et al., 2010; Leach et al., 2000; Leach et al., 1999). This inequality results in many challenges related to water access, water quality, and wastewater treatment for these underserved communities. According to the Center for Public Integrity (CPI), approximately 30% of colonia residents in the US fail to obtain clean water (CPI, 2017). A Rural Community Assistance Partnership (RCAP, 2015) report further detailed that an estimated 30% of the 840,000 colonia residents lack access to safe drinking water. As a result, people in colonias often suffer from severe health problems, such as waterborne diseases (Arsenault, 2021; Wilson et al., 2021).

To address this urgent issue, researchers from the social and environmental sciences, hydrological engineering, law and policy, and data science have begun to investigate feasible solutions leveraging existing and new social, physical, and cyber infrastructures to reduce water problems and increase clean water access and security in colonia communities (Wutich et al., 2022; Li et al., 2019). For instance, interviews and questionnaires have become popular approaches to gaining an understanding of water accessibility and security from a qualitative perspective (Arsenault, 2021; Garcia et al., 2016; Wilson et al., 2021). Araya et al. (2019) conducted a survey involving 92 residents of a non-border colonia in Central Texas and found that more than half of the respondents referred to an alternative drinking water source, while one third had safety concerns over the tap water in their system. However, the interview process is often labor and time intensive, making it difficult to extend and support large-scale analysis. Other studies have used a geographic information system (GIS) to analyze and map water conditions in colonias, providing an intuitive approach to examining the water access and health issues of these residents (Parcher & Humberson, 2009). While automation can be achieved, the criteria for classifying colonias in terms of their water poverty or level of water access often need to be manually defined (Korc & Ford, 2013).

In recent years, researchers have started to apply machine learning techniques to reveal water access and insecurity issues (Li, 2020; Zhong et al., 2021). Compared to traditional approaches, e.g., linear regression, which are better at handling small, high-quality datasets, machine learning have the ability to automatically mine complex big data to derive new information, patterns, and insights leveraging data-driven discovery ( Li & Arundel, 2022). For instance, Orak et al. (2020) utilized a fuzzy C-means clustering method to evaluate the surface water quality of the Ergene River in Turkey. Anbari et al. (2017) applied Bayesian statistics to identify potential risk areas to inspect in wastewater systems. Additionally, random forests, an improved radial basis neural network, and K-means clustering methods have been employed to estimate the groundwater level in aquifers, which is a key factor related to the assessment of water availability (Banadkooki et al., 2020; Ben Salem et al., 2018; Kardan Moghaddam et al., 2021; Majumdar et al., 2020).

Although many machine learning techniques have been applied to water-related research (Shen et al., 2021), quantitative studies on water accessibility in underserved communities are rarely conducted. In particular, the water research community has paid little attention to the water access challenges facing colonia communities. Addressing these domain-specific problems does, however, require an interdisciplinary approach that integrates new research paradigms, such as data-driven analysis. In this paper, we utilize machine learning—specifically, clustering and classification analysis—to better understand water insecurity in terms of access to public water and wastewater treatment systems, water quality, and potential water hazards in underserved colonias. Our research will inform decision-making to allow the limited resources to be allocated to the communities most in need of improvements. Mining the nationwide colonia dataset will also enable us to categorize communities that face similar levels of water shortage and quality issues, and to identify factors that are causing water inequality problems. This way, we can better understand the water access problem and facilitate more science-based decision-making among local governments.

The remainder of the paper is organized as follows: Section 2 provides a review of the literature, with a focus on assessing the state-of-the-art methods and findings in existing work related to characterizing and understanding water insecurity issues in general and colonias in particular. Section 3 describes the study area and the colonia dataset in detail. Section 4 introduces our methodological pipeline, which combines clustering and classification analysis to facilitate an in-depth understanding of water insecurity issues in colonias. Section 5 discusses the policy implications of our findings and explains how the findings can complement existing expert-based priority rankings of infrastructure and other resource needs. Section 6 concludes the paper and discusses future research directions.

## 2. Literature review

Water insecurity assessment plays an important role in effective and intelligent water governance and water priorities. Previous work focused on surveys, questionnaires, and reports (Araya et al., 2019; Danielaini et al., 2019) to understand key factors influencing water insecurity. Pearson et al. (2021) surveyed the interrelationship between water insecurity, demographic information, and area and personal conflicts over water in Africa. They developed five questions related to water insecurity regarding the quantity, quality, reliability, accessibility, and cost of water resources and provided only two response options for each question (never or always). The author found that interpersonal conflict has a positive correlation with water insecurity and inaccessibility. Although this approach helps reveal residents' concerns about water insecurity and potential causes of water conflict, the derived answers (e.g., never or always) are brief, limiting one's understanding of underlying concerns of water insecurity. The survey, due to its time-consuming nature, is limited to a few communities and cannot be easily applied to a large-scale study.

There are also water insecurity investigations on colonias on the north side of the US-Mexico border, where the water conditions are highly concerning, and basic infrastructure is lacking. Wutich et al. (2022) reviewed the current research on water insecurity in the colonias, especially the governments' efforts to fund and support water research, the colonias' current water quality and accessibility conditions, and the cause of water insecurity. Based on the literature, they also proposed ways to improve accessibility to clean water, for instance, through building a coupled social and physical infrastructure, as well as policy reform. Jepson (2014) discussed water insecurity issues in the colonias and classified colonias' water access conditions into four categories: safe water access, marginally safe water access, marginally unsecured water access, and unsecured water access. Although this research offers a general picture of water access conditions, the method was qualitative and experience-based, with limited data applied to reflect the real-world conditions. Other researchers (e.g., Tippin et al., 2021; Zheng et al., 2022) also studied water security in the colonias, and similarly used a qualitative approach, such as summarizing literature and government reports to discuss the impact of water insecurity, potential water treatment, and water policies and laws that can be applied in the colonias. However, without actual data support, a comprehensive picture cannot be rendered regarding different water access challenges from which the colonias suffer.

With the emergence of big data and because of the intertwined factors influencing water security, including urbanization, climate change, population growth, and infrastructure construction (Astaraie-Imani et al., 2012; Doeffinger & Hall, 2021; Jaramillo & Nazemi, 2018), advanced quantitative approaches that can incorporate multiple factors in the analysis have become popular to help understand such intractable interactions among people, water, and infrastructures (Wheater & Gober, 2013). In particular, clustering methods have been extensively applied to study water accessibility, water quality, and water insecurity issues (Bilgin, 2018; Celestino et al., 2018; Cooper-Vince et al., 2018; Krishnaraj & Deka, 2020). K-means clustering, one of the most popular clustering methods, has been applied to investigate the chemical composition of groundwater systems (Zhang & Gu, 2014; Celestino et al., 2018), assess community accessibility to clean water resources (Radliya et al., 2019), and identify different zones based on water quality measures (Mandel et al., 2015). Other clustering approaches— such as fuzzy C-means clustering, Getis-Ord $G_i^*$ clustering, and Bayesian clustering—have been used to investigate the spatial distribution of issues related to water quality, such as waterborne diseases (Altherr et al. 2019) and contaminant mixtures (Hoover et al., 2018).

While clustering techniques are popular and helpful for identifying similar water access conditions across communities, they also suffer from some inherent limitations. First, researchers must manually (in K-means clustering and other similar techniques) define parameters, such as the number of clusters, which requires knowledge of the distribution pattern within the data. Second, results of the clustering analysis are not self-explanatory, making it difficult to interpret how data points are grouped together, especially when multiple attributes are used. Third, traditional clustering algorithms can only handle numerical data as input, which poses a challenge when data attributes are of mixed types, including numerical and categorical. Fourth, existing clustering analyses often ignore the analysis of the importance of factors that distinguish each group's water access conditions from the others, impeding the derivation of distinctive characteristics of each cluster and achieve in-depth analysis. Consequently, increasing the explainability of clustering algorithms is valuable to ensure the trustworthiness of the results and any policy decisions made based on them (Goodchild & Li, 2021).

To address the aforementioned challenges, in this paper, we aim to apply a data-driven analysis that combines clustering with a decision tree-based machine learning approach. Our central goal is to understand the water access conditions in the 2000+ colonias within the US, the living conditions of which are very concerning, but their water access challenges have not been well studied. To the best of our knowledge, there has been no comprehensive analysis of water access conditions across all colonias. The proposed clustering approach helps derive different levels of water insecurity challenges from which colonias communities suffer. The decision tree-based classification was further applied to analyze the results of clustering to rank the importance of the factors causing this inequality. This latter part of the analysis has rarely been applied in relevant studies; hence, it remains unclear how differently these factors affect the colonias. This data-driven analysis framework will help render a complete picture of the priority of factors affecting water security in the colonias, such that targeted improvement plans can be developed and applied to treat the most urgent and significant factors in different colonias groups. The next section describes our methodology in detail.

## 3. Study areas and datasets

The datasets used in this study were collected from RCAP, a nonprofit network of technical assistance providers, which in 2015 collected comprehensive information about water availability for a total of 2190 colonias from 35 counties within the 4 US–Mexico border states (i.e., California, Arizona, New Mexico, and Texas). Specifically, there are 107 named colonias in Arizona, 35 in California, 158 in New Mexico, and 1890 in Texas. Figure 1 illustrates the geographical distribution of these colonias.

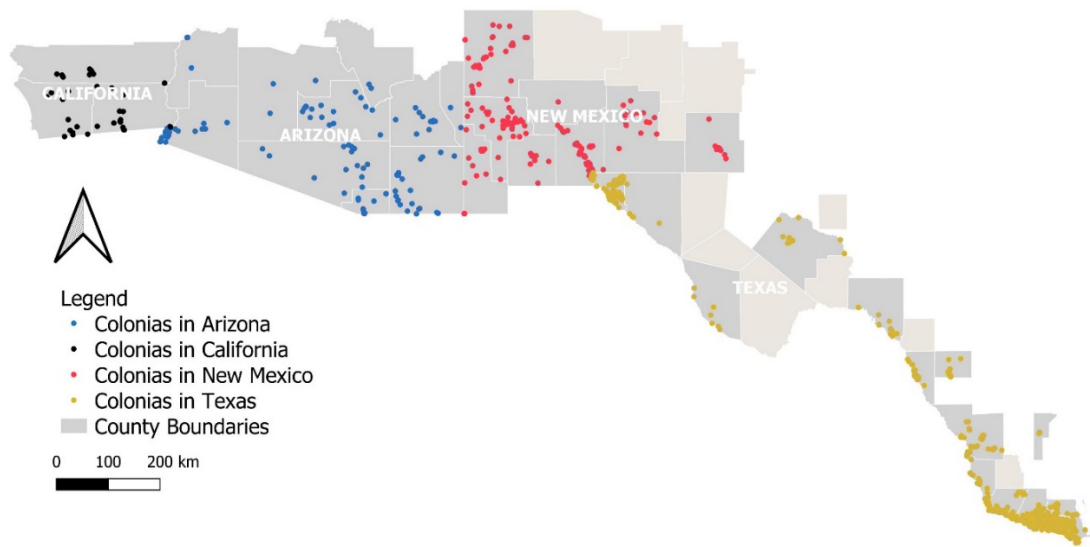

**Figure 1.** Geographical distribution of colonias in the study area. Dots show the colonia locations. Polygonal areas indicate county boundaries. Tan polygons indicate counties that do not contain colonias but are depicted for the purpose of drawing a continuous study area. CA: California, AZ: Arizona, NM: New Mexico, TX: Texas.

To analyze the severity of the water insecurity in colonias, we utilized 12 available attributes describing the water supply, water sanitation, and water quality in the communities. The attributes contain both numerical and categorical data types. The descriptions of these attributes provided in Table 1 are quoted from the RCAP Colonias Phase II Assessment Report (RCAP, 2015).

**Table 1.** Description of the selected attributes.

| Data Type | Attribute | Description |
|---|---|---|
| **Categorical** | Water Source | Description of water source(s) serving the colonia |
| | Water Hauled | Are residents of the colonia hauling water? (Yes/No) |
| | Private Wells | Are residents of the colonia served by private wells? (Yes/No) |
| | Public Water Service | Are residents of the colonia served by a public water system? (Yes/No) |
| | Service Adequacy | Is service provided to this colonia by a public water system adequate? (Yes/No/Partial) |
| | Water Health Hazard | In terms of water supply service in this colonia, is a health hazard indicated? (Yes/No) |
| | Served by Public Sewer | Is the colonia served by a public wastewater disposal system? (Yes/No) |
| | Estimated Population | Estimated population of the colonia |
| **Numerical** | People without Water | Provide the estimated number of persons in this colonia that are NOT served by a public water system |
| | People without Wastewater | Provide the estimated number of persons in this colonia that are NOT served by a public wastewater system. |
| | People with Water | Provide the estimated number of persons in this colonia that ARE served by a public water system. |
| | People with Wastewater | Provide the estimated number of persons in this colonia that ARE served by a public wastewater system |

## 4. Methodology

In this research, a new machine learning pipeline was developed to combine both clustering and classification analysis to help understand the water access insecurity issues facing colonias. This pipeline addresses four major challenges facing existing approaches: (1) incorporating complex data types; (2) achieving (near) automation in the process; (3) increasing the explainability of the model's learning process; and (4) combining expert knowledge (i.e., top-down) with data-driven (i.e., bottom-up) approaches to derive meaningful results.

Figure 2 demonstrates the entire workflow. We first introduced a hierarchical clustering strategy to divide the colonias into two initial groups: those with and without public water services. The reason for adopting top-down hierarchical clustering and incorporating prior knowledge was to achieve both efficiency and accuracy in the analysis (Takumi et al., 2012). As the relevant literature revealed, the presence of a public water supply system is one of the most important measures for water accessibility (Hall et al., 2014; Nathanson, 2020), as it provides a source of drinking water that is safer than that of private wells, the water quality of which is not regulated by the Environmental Protection Agency (EPA; RCAP, 2015; Law et al., 2017). Accordingly, grouping the colonias into these two broad classes, as discussed above, can help in identifying and understanding the prominent and unique characteristics of colonias with different water access conditions. In terms of efficiency, putting all the data points into the clustering analysis is more computationally expensive than is leveraging hierarchical clustering. Hierarchical clustering is also effective in handling complex data with multiple attributes. When all data points are simultaneously considered in the clustering process, the algorithm may get "confused" and create sub-optimal results that are more difficult to explain. Using the strategy of hierarchical clustering, the most similar data points are partitioned into logically meaningful clusters (i.e., subsets); then, further analysis only requires working on each subset, reducing the computation cost through this divide-and-conquer strategy.

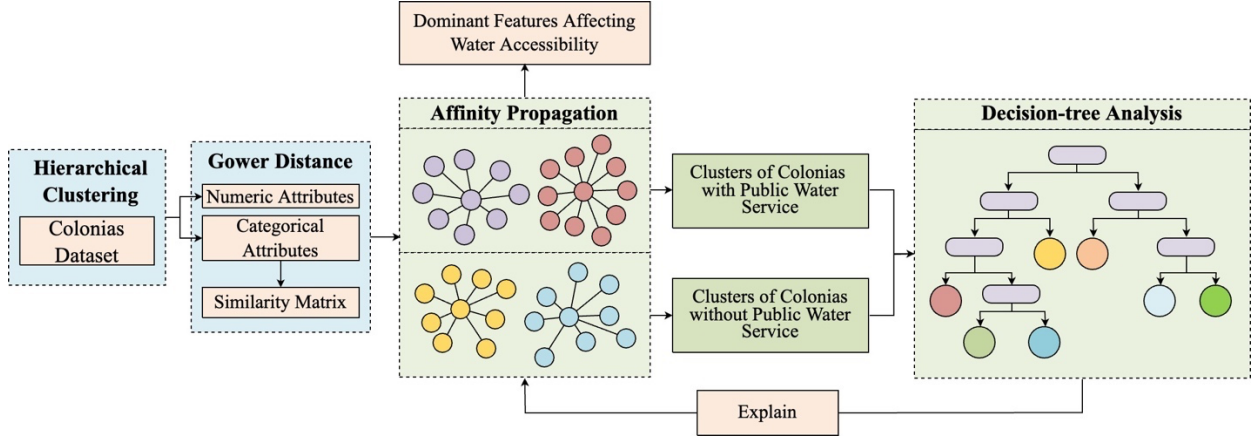

**Figure 2**. Workflow of the proposed methodology.

Once the initial data partition was completed, the clustering algorithm was applied to each subset of the colonias. Here, we applied the APA (Frey & Dueck, 2007) in our clustering analysis. Compared to other clustering algorithms, the APA does not require a user-defined cluster number as the input parameter, allowing more flexibility and autonomy in the exploration process. The algorithm also introduces a message-passing mechanism to determine an optimal cluster assignment, which has proven to be effective in a variety of applications, such as the clustering of images, natural language processing, and gene detection (Frey & Dueck, 2007). To apply the APA, we first utilized the Gower distance to calculate the similarity between each pair of colonias considering all the attributes, both numerical and categorical. The result was a similarity matrix that served as the input for the clustering algorithm. Once more clusters were generated, we used the cluster label as the ground truth to further apply a decision tree analysis, to derive the most prominent and unique characteristics of each cluster. This way, we could gain a deeper understanding of the machine learning process and the water access challenges faced by different groups of colonias.

*4.1 Hierarchical clustering*

Following the principle of hierarchical clustering, the 2190 included colonias were initially grouped into 2 clusters, based on the categorical attribute, "has public water service." Table 2 shows how many colonias fall into each initial cluster and their distribution by state. As can be seen, the majority of colonias in California (i.e., more than 97%) has public water supply services, meaning that the living conditions of colonias in California are, in this regard, the best among all the states. This is followed by Texas and Arizona. New Mexico has the lowest rate of water accessibility, with only 60% of the colonia communities having access to a public water service.

**Table 2**. Colonia statistics for each subset by state

| Subset | Arizona | California | New Mexico | Texas | Total |
|---|---|---|---|---|---|
| **Colonias with public water service** | 84 | 34 | 95 | 1726 | 1939 |
| **Colonias without public water service** | 18 | 1 | 63 | 135 | 217 |
| **Percentage of colonias with public water service** | 82% | 97% | 60% | 93% | 90% |

Before sending the datasets for further clustering, we performed a normalization of the numerical attributes, including the number of "people without water," "people without wastewater," "people with water," and "people with wastewater" by "estimated population" (the descriptions of which can be found in Table 1), to eliminate the scale effect caused by different sizes of colonias. Further, colonias with abnormal normalized values (i.e., those greater than 1) were considered noise and removed from the analysis. This resulted in 1939 colonias with public water services and 217 colonias without public water services. The next section explains how the similarity among the colonias was computed, considering multiple numerical and categorical attributes.

*4.2 Similarity Measure based on Gower Distance*

Clustering methods rely on a similarity measure to determine the assignment of data points into clusters. Generally speaking, similar data points will be grouped in the same cluster and dissimilar data points will be assigned to different clusters. Euclidean distance is a popular way to derive the (dis)similarity between data points, but it can be applied only to attributes with numerical values. Since the colonias dataset contains both numerical and categorical attributes, there is a need to quantitatively define the similarity values before applying the clustering algorithm. In this paper, we adopted the Gower distance (Gower 1987) to determine the distance (i.e., dissimilarity) between data points of mixed types. Mathematically, the Gower distance $d_{ij}$ between data points $i$ and $j$ can be defined as:

$$d_{ij} = \frac{1}{N} \sum_{k=1}^{N} dis_{ij}^k$$

(1)

, where $dis_{ij}^k$ is the dissimilarity between data points $i$ and $j$ on attribute $k$, and $N$ is the total number of attributes. For numerical attributes, the dissimilarity value $dis_{ij}^k$ is based on Manhattan distance (Gower 1987), which can be expressed as:

$$dis_{ij}^k = |x_i - x_j|/R_k$$

(2)

, where $R_k$ is the value range of attribute $k$. $x_i$ and $x_j$ are the value of data points $i$ and $j$ on attribute $k$, respectively. $|x_i - x_j|/R_k$ is the normalized Manhattan distance. The smaller the $dis_{ij}^k$ is, the more similar the data points are in terms of attribute $k$. When the two data points are the same, the dissimilarity value is 0.

For categorical attributes, the dissimilarity calculation is based on the Dice distance, which calculates the total number of matches in different attributes for two data points (Mulekar et al. 2017). Before applying them to compute the Dice distance, we converted the categorical attributes into the dummy variables so that each categorial attribute and its corresponding values will be converted to multiple binary attributes with the value being 1 or 0. This process is also called one-hot encoding, a popular approach to transform categorical attributes to use in a machine learning process (Brownlee 2017). Here we chose the attribute "*Private Wells*" for colonias Pleasanton and Wilco as an example. The "*Private Wells*" has two values, namely "Y" (Yes) and "N" (No), as Table 3 shows. After converting it into a dummy variable, the new attribute will be named by the original name followed by its value choice, namely "Private Wells_Y" and "Private Wells_N". Because colonia Pleasanton does have private wells, so the value for "Private Wells_Y" for it is 1 and "Private Wells_N" is 0, as shown in the Table 4.

Table 3. Colonias and their "private wells" attribute

| Colonia Name | Private Wells |
|---|---|
| Pleasanton | Y |
| Wilco | N |

Table 4. Value assignment after conversion to a dummy attribute

| Colonia Name | Private Wells_Y | Private Wells_N |
|---|---|---|
| Pleasanton | 1 | 0 |
| Wilco | 0 | 1 |

Equation (3) provides the formula for calculating the dissimilarity (or distance) of two data points based on its categorical attributes and dice distance. Again, the smaller the distance between two points is, the more similar they are. If the values of two data points are the same, their dice distance is 0.

$$dis_{ij}^k = \frac{N_{FP} + N_{FN}}{2N_{TP} + N_{FP} + N_{FN}}$$

(3)

In the formula, $N$ provides the count of converted dummy attributes from the original attribute $k$ that yield a specific value combination for data points $i$ and $j$. For instance, $FP$ means false positive, so any value pair yielding <0,1> or <1,0> will fall in this category. $TP$ means true positive, indicating the case when the variable values are true (1) for both data points. $FN$ means false negative; it refers to when the attribute values are both 0s for points $i$ and $j$. Given this measure, the dice distance between the two colonias Pleasanton and Wilco on attribute "private wells" described in Table 4 will be 1, given the very different values they have for the converted variables. Similar to the normalized Manhattan distance given in Eq (2), the dice distance also has a value range of [0,1].

Compared to other similarity measures, such as the similarity based on the Euclidian distance and Manhattan distance, which work only for numerical values, the Gower distance combines different dissimilarity calculations for both numerical attributes and categorical attributes. And these values are given equal weight such that the final similarity result will not be biased toward one or another. Therefore, it is more suitable for a complex dataset which contains mixed types of attributes. In the next section, we discuss the creation of a similarity matrix leveraging the Gower distance.

*4.3 Creation of a similarity matrix*
One required input of the affinity propagation algorithm is an $N * N$ similarity matrix providing the similarity between each pair of any two data points (there are in total $N$ data points) in the dataset. This matrix could be symmetric, when the similarity between data points $i$ and $j$ and $j$ and $i$ are the same. The matrix could also be dissymmetric according to different application needs. In our study, a symmetric similarity matrix is generated based on the Gower distance, a measure of dissimilarity discussed above. The similarity value between data point $i$ and $j$ is denoted as $s_{ij}$.

As for the similarity measure based on the Gower distance $d_{ij}$, the smaller the Gower distance is, the more similar two data points are. We need to properly convert the Gower distance into similarity values, with a larger value indicating higher similarity between two points. Because the affinity propagation algorithm requires all similarity values to be negative, we introduced a negative parameter $\theta$ to convert the Gower distance into similarity values. The equation is shown below:

$$s_{ij} \leftarrow \theta * d_{ij}, (\theta < 0)$$

(4)

Here we adopt -1 as $\theta$'s value, such that the product of $\theta$ and $d_{ij}$ will always be a negative value between -1 and 0, and this also meets the requirement that the smaller the distance is, the larger $s_{ij}$ is. One special treatment is for the self-similarity between the $i$-th data point with itself, namely $s_{ii}$. Instead of relying on the above equation, these self-similarity values can be customized, as they give a sense of preference (higher value) for a data point $i$ to serve as a potential cluster center (a.k.a., exemplar). If no preference is given for any data point, the median value in the similarity matrix will be set as the preference for each $s_{ii}$ before starting the clustering process.

### 4.4 Affinity propagation clustering

Once the similarity matrix is created, it can be used as the input of the affinity propagation algorithm as the criterion for data assignment. As discussed, an APA is capable of grouping data points without asking users to provide a predefined number of clusters. The goal of the affinity propagation algorithm is to find the exemplar point for each cluster and the assignment of all other points to one of the exemplars. It achieves this by developing a novel message passing mechanism.

Specifically, all the data points are considered as nodes; and messages can be passed both ways between any two data points, forming a message passing network (Figure 3). The goal of message passing is to identify exemplar data points for each cluster based on which other data points will choose to join or to be included by that cluster. Initially, any data point could be an exemplar. The directed arrows in Figure 3 (Frey and Dueck 2007) represent the direction through which a message ($r_{ik}$ or $a_{ik'}$, which will be introduced in the next section) is passing between a data point $i$ and the candidate exemplars $k$ and $k'$.

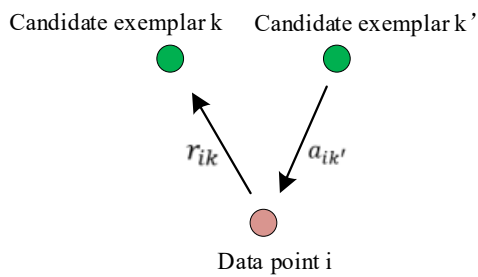

**Figure 3**. Messages passing among the data points to cluster.

After loading the similarity matrix into the affinity propagation method, two types of messages will be passed through the edges for information exchange: responsibility ($r_{ik}$) and availability ($a_{ik'}$). First, the responsibility value $r_{ik}$ going from the $i$-th to $k$-th data point indicates how well the $k$-th data point is suitable for being the exemplar of $i$-th data point. In each iteration, the updated responsibility value is formulated in (5) where $s_{ik}$ is the similarity between data point $i$ and $k$, and $a_{ik'}$ represents the availability score of another candidate exemplar $k'$ for the $i$-th point to choose it as the exemplar.

$$r_{ik} \leftarrow s_{ik} - \max_{k' \neq k}\{a_{ik'} + s_{ik'}\}$$

(5)

This equation infers that if another candidate exemplar, the $k'$th point has higher availability and is more similar to the $i$-th point (a high $\max_{k' \neq k}\{a_{ik'} + s_{ik'}\}$ value), then candidate $k$'s suitability for attracting the $i$-th point into its cluster will be reduced (reflected by a lower $r_{ik}$).

Next, the availability value $a_{ik}$ passing from the $k$-th to $i$-th data point can be calculated using Equation (6):

$$a_{ik} \leftarrow \min\left\{0, r_{kk} + \sum_{i' \notin \{i,k\}} \max\{0, r_{i'k}\}\right\}, i \neq k$$

(6)

, where the max{} function removes some negative bias from the $i'$th point for not selecting $k$ as an exemplar (when a lower $r_{i'k}$ value is present), because point $i'$ could be very far away from $k$ and may reasonably prefer another candidate exemplar. So, preference for the exemplar ($r_{i'k}$) of such a point $i'$ could be very much biased. Therefore, such negative biases $r_{i'k}$ are removed and not considered for calculating $a_{ik}$. On the other hand, the min{} function removes some strong positive bias from $k$'s neighbors to select $k$ as an exemplar. $r_{kk}$ is the preference value for the $k$-th point to become an exemplar. By combining the preference of selecting itself as an exemplar as well as other points' interest in joining $k$, we can derive the availability of $k$ to become a cluster exemplar for point $i$.

For the $k$-th data point itself, the availability $a_{kk}$ is updated according to Equation (7):

$$a_{kk} \leftarrow \sum_{i' \notin k} \max\{0, r_{i'k}\}$$

(7)

During the clustering process, the availability $\{a_{ik}\}$ and responsibility matrices $\{r_{ik}\}$ will be updated iteratively. The objective is to find the set of {k} that maximizes $\{a_{ik}+r_{ik}\}$. The process will stop when the summation matrix $\{a_{ik}+ r_{ik}\}$ becomes stable and its values are unchanged. For each data point $i$ (i-th row), the column index $k$ where the largest value of that row is found is an exemplar. The cell ($i,k$) therefore indicates the assignment of point $i$ to the cluster represented by exemplar $k$. This way, the cluster assignment can be automatically accomplished.

*4.5 Adaptative affinity propagation to address the oscillation issue*

Although the affinity propagation algorithm has the merit of generating clusters without a redefined K (number of clusters), it still suffers from the oscillation issue (Refianti et al., 2016; Wang et al., 2007). Oscillation happens when the exemplar keeps changing at each iteration, making the algorithm difficult to converge. To solve this problem, a damping factor $\gamma$ (0.5 <= $\gamma$<1) is introduced to reduce the oscillation and increase the algorithm's adaptiveness. At each iteration, the current messages (i.e., the responsibility and availability values) will be determined by the weighted sum of the values in both current and previous iterations. Equations (8) and (9) provide the rules for updating the responsibility ($r_{ik}^{t+1}$) and availability ($a_{ik}^{t+1}$) values at the t+1 iteration with the damping factor.

$$r_{ik}^{t+1} \leftarrow \gamma * r_{ik}^{t} + (1 - \gamma) * r_{ik}^{t+1} \tag{8}$$

$$a_{ik}^{t+1} \leftarrow \gamma * a_{ik}^{t} + (1 - \gamma) * a_{ik}^{t+1} \tag{9}$$

By carrying information from the previous iteration, the algorithm will avoid sharp turns in searching for the optimality in the solution space, making it easier and faster to converge.

*4.6 Evaluation*

After employing the affinity propagation algorithm, we also need to evaluate the clustering results. A Silhouette score (SS) provides a measure of how good the clustering algorithm has separated the data points into meaningful clusters. It is calculated by using the average $d_1$ of the distances between any point pair within all the clusters and the average $d_2$ of the distances between any point pair falling in different clusters. The formula can be expressed as:

$$SS = \frac{(d_2 - d_1)}{\max(d_1, d_2)} \tag{10}$$

The Silhouette score is high when a clustering algorithm has well partitioned the points based on their similarity (or inversed distance), in which case $d_2$ will be large enough and $d_1$ is small enough. The Silhouette score reaches its upper limit (1) when $d_1 \ll d_2$. This indicates very good clustering performance. The score reaches its lower limit (-1) when $d_1 \gg d_2$, when the algorithm does poorly in grouping similar data points.

5. **Results**

*5.1 Clustering result analysis*

We utilized the scikit-learn 0.21.3 python package (Python version 3.8) to apply the clustering task to our selected colonia datasets. For the clustering analysis of the two initial groups (i.e., those with and without public water services), we selected a series of damping factors within the range of [0.5, 0.9] (recommended by Hu et al. [2019] and Kramer [2016]) and an interval of 0.1 to evaluate the clustering results using the Silhouette score. The change in Silhouette scores for each subset with changing damping values is shown in Figure 4. It can be observed that a damping factor of 0.6 results in the highest Silhouette score for both datasets.

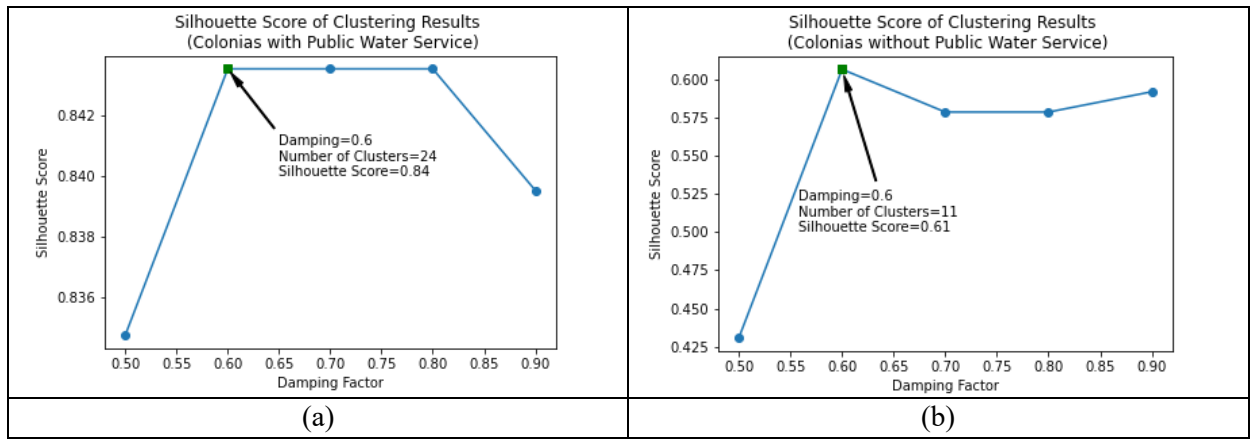

**Figure 4** Clustering performance with changing damping factor values for colonia subsets (a) with a public water service and (b) without a public water service.

Based on this analysis, we chose the best clustering results with the highest Silhouette score for each colonia subset. For the 1937 colonias with public water services, 24 clusters were generated. Among these, the top 10 largest clusters contained more than 97% (1878) of the original dataset. These 10 clusters are listed in Table 5. For the 217 colonias without public water services, 11 clusters were generated. Detailed statistics describing the colonias in each cluster in these two subsets are presented in tables 5 and 6.

**Table 5**. Number of colonias with public water services in each cluster across the four states (top 10 clusters).

| Cluster | Arizona | California | New Mexico | Texas | Total |
| --- | --- | --- | --- | --- | --- |
| **Cluster 21** | 10 | 0 | 14 | 836 | 860 |
| **Cluster 19** | 18 | 14 | 20 | 805 | 857 |
| **Cluster 11** | 10 | 0 | 23 | 12 | 45 |
| **Cluster 9** | 4 | 1 | 17 | 7 | 29 |
| **Cluster 15** | 0 | 0 | 0 | 20 | 20 |
| **Cluster 4** | 1 | 16 | 0 | 0 | 17 |
| **Cluster 5** | 8 | 0 | 6 | 2 | 16 |
| **Cluster 20** | 5 | 0 | 0 | 8 | 13 |
| **Cluster 22** | 0 | 0 | 0 | 12 | 12 |
| **Cluster 1** | 8 | 0 | 0 | 1 | 9 |
| **Total** | 64 | 31 | 80 | 1703 | 1878 |

From Table 5, it is evident that clusters 21 and 19 account for a larger number of colonias than do the others. They comprise more than 89% of the total colonias in this dataset and are mainly located in Texas. In Arizona, the colonias are widely spread across almost all of the top 10 clusters, except for clusters 15 and 22. Colonias in California exist only in clusters 19, 9, and 4. Moreover, clusters 19 and 4 include the highest proportion of colonias in California. No colonias in New Mexico are found in clusters 15, 4, 20, 22, or 1. Approximately 96% of the colonias in Texas appear in clusters 21 and 19.

**Table 6.** Number of colonias without public water services in each cluster across the four states.

| Cluster | Arizona | California | New Mexico | Texas | Total |
|---|---|---|---|---|---|
| **Cluster 3** | 7 | 0 | 15 | 40 | 62 |
| **Cluster 0** | 2 | 0 | 43 | 16 | 61 |
| **Cluster 10** | 0 | 0 | 2 | 54 | 56 |
| **Cluster 7** | 0 | 0 | 2 | 9 | 11 |
| **Cluster 2** | 5 | 0 | 0 | 2 | 7 |
| **Cluster 6** | 1 | 0 | 0 | 4 | 5 |
| **Cluster 13** | 0 | 0 | 0 | 4 | 4 |
| **Cluster 8** | 0 | 0 | 0 | 4 | 4 |
| **Cluster 14** | 3 | 0 | 0 | 0 | 3 |
| **Cluster 12** | 0 | 0 | 1 | 2 | 3 |
| **Cluster 16** | 0 | 1 | 0 | 0 | 1 |
| **Total** | 18 | 1 | 63 | 135 | 217 |

Table 6 shows the clustering results for colonias without public water services. With 217 of such colonias in total, there is only 1 in California, and it exists in cluster 16. Colonias in New Mexico appear in clusters 3, 0, 10, 7, and 12. Colonias in Arizona are mainly situated in clusters 3, 0, 2, 6, and 14. In addition, colonias in Texas appear in all clusters but 14 and 16.

*5.2 Decision tree analysis and results*

After the clustering process, all data points were assigned to a cluster. Besides this exploratory analysis, we went one step further, mining the clustering results to derive the prominent characteristics of the colonias in each cluster and to identify the dominant water access insecurity challenges that differ from cluster to cluster. To achieve this goal, a decision tree analysis was applied to the clustering results. As a supervised learning method, decision tree analysis can be used for both classification and regression. It recursively divides datasets into smaller subsets based on different value or choice conditions in the attributes until the partitioned result matches the ground truth. The resultant decision tree will provide multiple decision paths to show how the data partition is done. In our case, it shows how the clustering results are generated. The nodes (i.e., attributes) near the roots of a decision tree are more important than those located near the leaves of the tree. This can also indicate the importance of an attribute in determining the final partitioning/clustering results.

There are various decision tree algorithms that can be applied; among these, the classification and regression tree (CART) algorithm supports numerical attributes during the decision-making process. Therefore, we chose to use the CART algorithm to implement the decision tree analysis. To run the CART model, we fed the colonia dataset along with the colonia attributes as the input, while the clustering results served as the ground-truth labels. Figure 5 shows the results of the analysis. Each rectangle in the figure refers to an attribute, and the branches coming out of it are the conditions derived when rendering different decisions (i.e., clusters). The oval shapes contain the final cluster results and the number of samples that fall in the decision path (i.e., cluster). There may be multiple paths that result in the same cluster, and these are indicated by the use of the same color at the leaf nodes. The root node is the attribute, "colonias with public water service," and it has two values (i.e., "yes" and "no"). All the colonias with a "yes" answer go to the upper subtree, and those with a "no" answer go to the lower subtree. As access to a public water system is critical to obtaining safe drinking water, all the colonias and clusters labeled in the upper subtree can be considered as having a better water access condition than those in the lower subtree. As the subtrees continue to split, more branches are created, and each branch eventually reaches a cluster group (indicated by leaf nodes, in an oval shape). The attributes and their values along the path (from the root node to the leaf node) showcase the shared water access conditions of the colonias in a cluster.

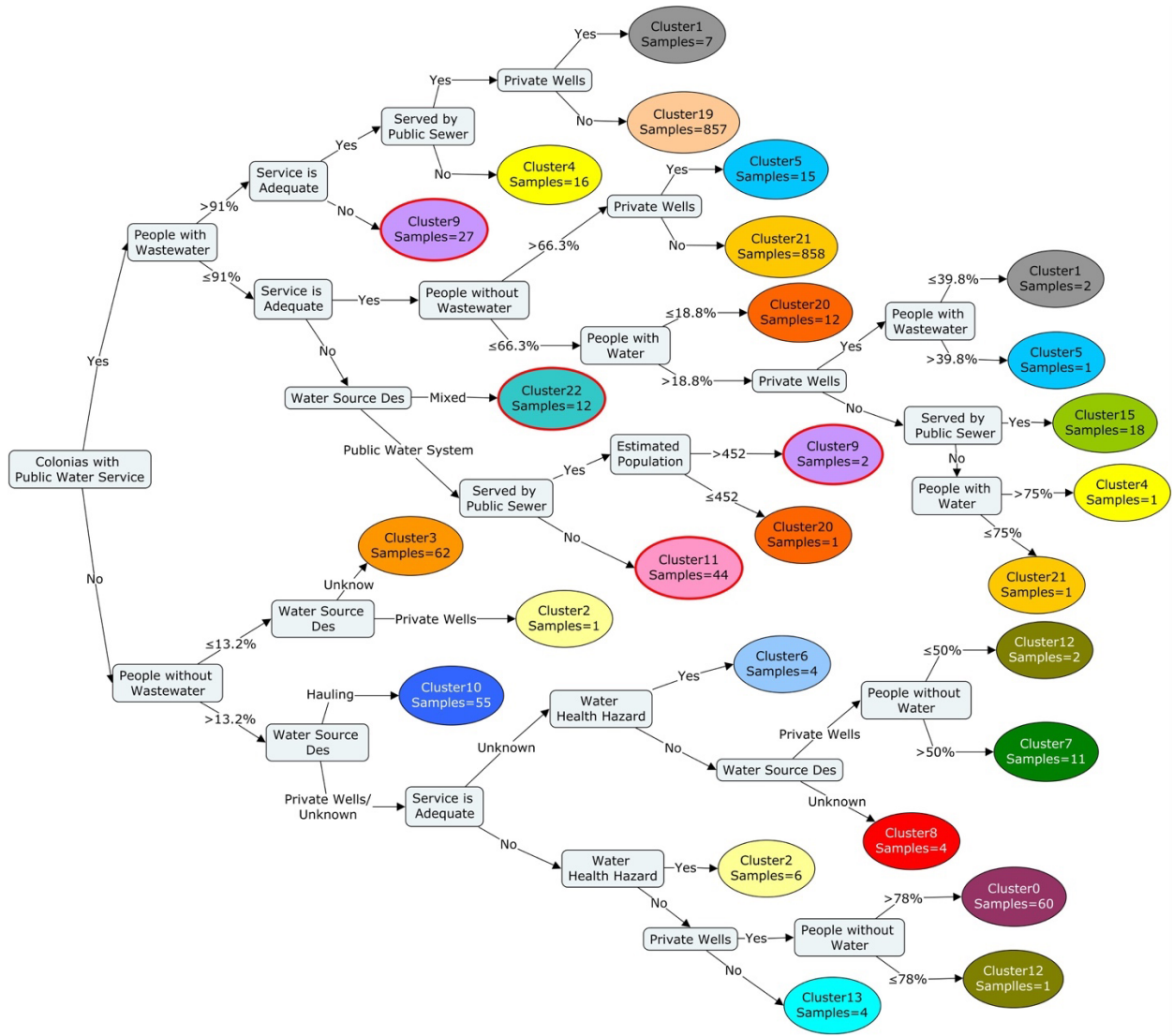

**Figure 5.** Decision tree analysis results for colonias with and without public water services.

Figure 6 lists important rules extracted from the decision tree that yielded the five largest clusters in the two largest subtrees (i.e., colonias with and without public water services). The sample number and percentage value in each oval shape provide the number and proportion of colonias that follow this rule in each cluster. The rule number also matches the cluster number. For instance, Rule 21 in Figure 6(a) characterizes colonias that have access to a public water system, in which no one relies on private wells, in which the service level is deemed to be adequate, but in which most residents do not have access to wastewater treatment services. More than 99% of the colonias in Cluster 21 satisfy all the conditions in this rule. Hence, it characterizes the common properties of the colonias and their water access conditions in this cluster. In contrast, colonias in Cluster 19 have better water access conditions, as they have access to both public water service and public sewer systems. Colonias in Cluster 15 (Rule 15) also have an adequate level of water service, although a proportion of these communities do not have access to wastewater treatment services.

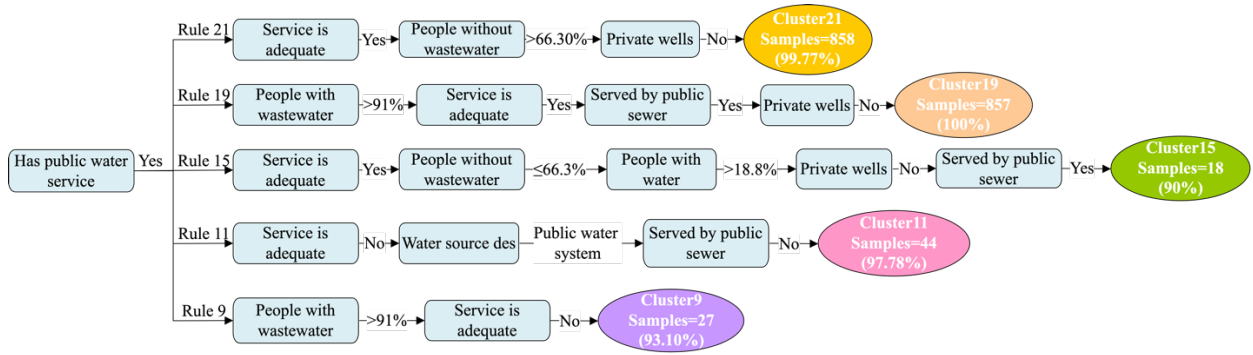

(a) Decision paths for the five largest colonia clusters with public water services. Clusters 21, 19, and 15 have adequate water services, while clusters 11 and 9 do not.

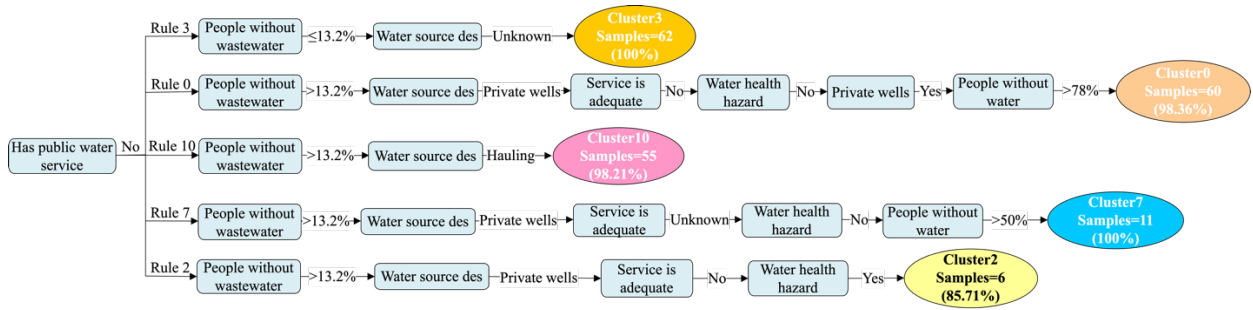

(b) Decision paths for the five largest clusters without public water services.

**Figure 6.** Top 5 decision paths for colonias falling in different clusters: (a) colonias with public water services, and (b) colonias without public water services.

As shown, the service level in colonias in Cluster 11 (see Rule 11 in Figure 6[a]) is not as high as that in colonias in Clusters 15, 19 and 21. Although they all have access to public water services, these communities are not connected to a public sewer system, so the overall service level for colonias in Cluster 11 is identified as inadequate. A surprising result is evident in Cluster 9 (Rule 9); although nearly all these colonias have access to public water services, and most of them (more than 91%) have wastewater treatment services, the service level in these communities is still identified as inadequate. Further investigation is needed to determine what caused the service level to be marked as insufficient.

For colonias in Cluster 3 (Rule 3), listed in Figure 6(b), although they do not have access to public water services, the majority of residents in these colonias do have access to wastewater treatment services (i.e., only 13.2% of residents do not). The water source for these colonias is not yet clear. For Cluster 0 (Rule 0), more than 78% of people do not have access to a public water supply; instead, private wells are their main water source. Further, at least 13% of people are not provided with wastewater treatment services. Although the service level is identified as inadequate, there is no known water health hazard. Cluster 7 (Rule 7) has a condition like that of Cluster 0, as private wells are the main water source and most people do not have access to either a public water service or a wastewater treatment system. The geographical distributions of the colonias in clusters 0 and 7 are shown in Figure 7(a).

The living conditions in Cluster 10 (Rule 10) are a bit more concerning, as the residents of these colonias still have to haul their water from another source. Most of these colonias are relatively small (i.e., with populations of fewer than 468 people). As Figure 7(b) shows, they are mostly located near the border of Texas and New Mexico, as well as in the southeast corner of Texas. Colonias in Cluster 2 are also suffering from severe water access problems; in addition to not having adequate services, there is a reported water health hazard, endangering the wellness of the colonia residents. Although there are only five such colonias in this group, the living conditions in these colonias require immediate attention. The distributions of colonias in these clusters are shown in Figure 7(c).

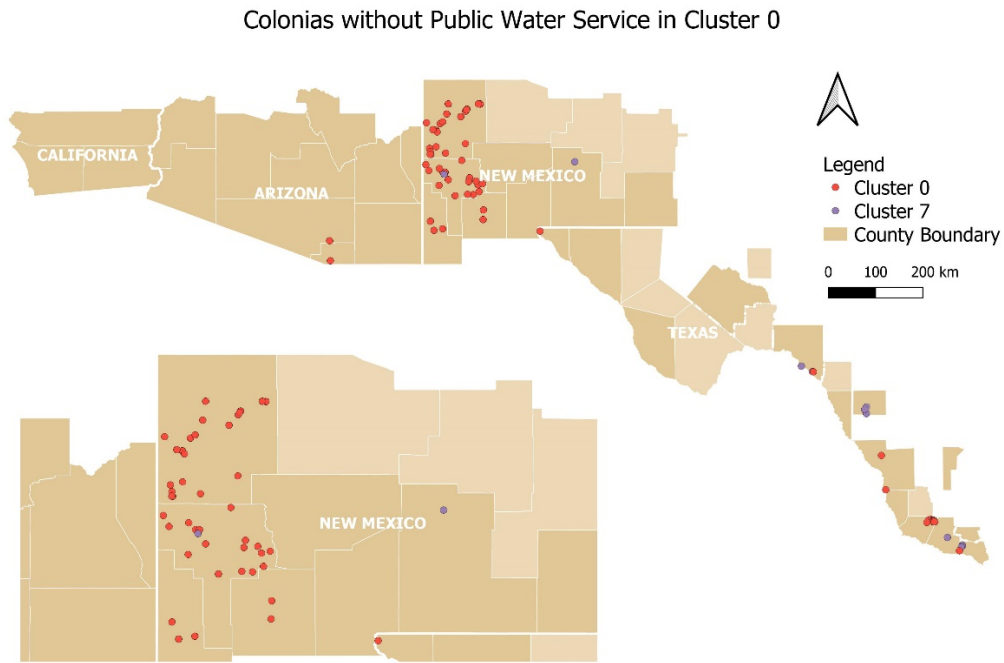

(a) Distribution of colonias in clusters 0 and 7, which have similar water access conditions.

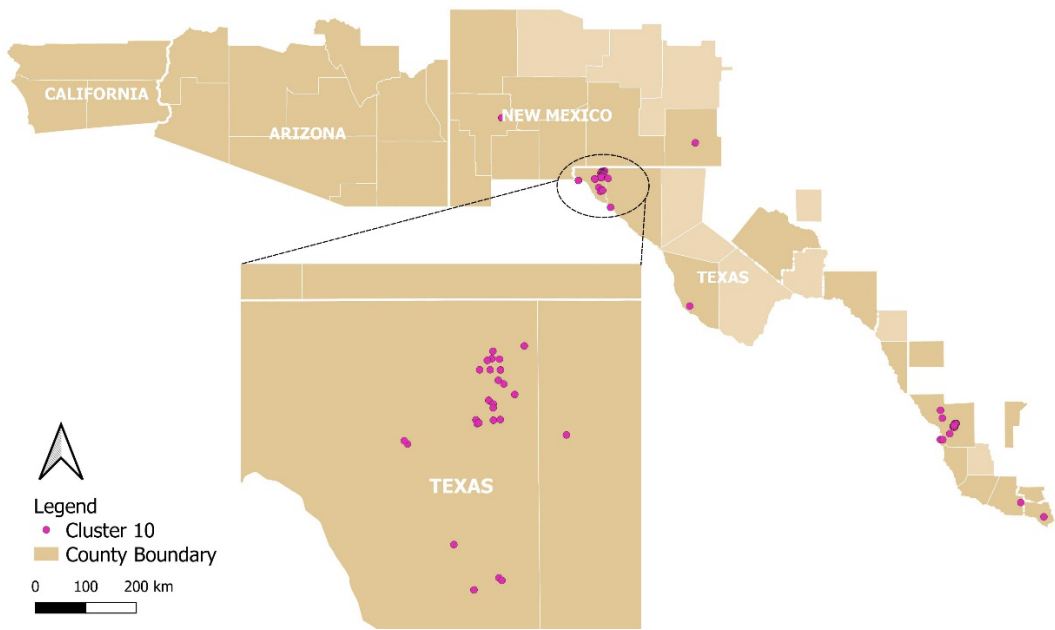

(b) Distribution of colonias in Cluster 10.

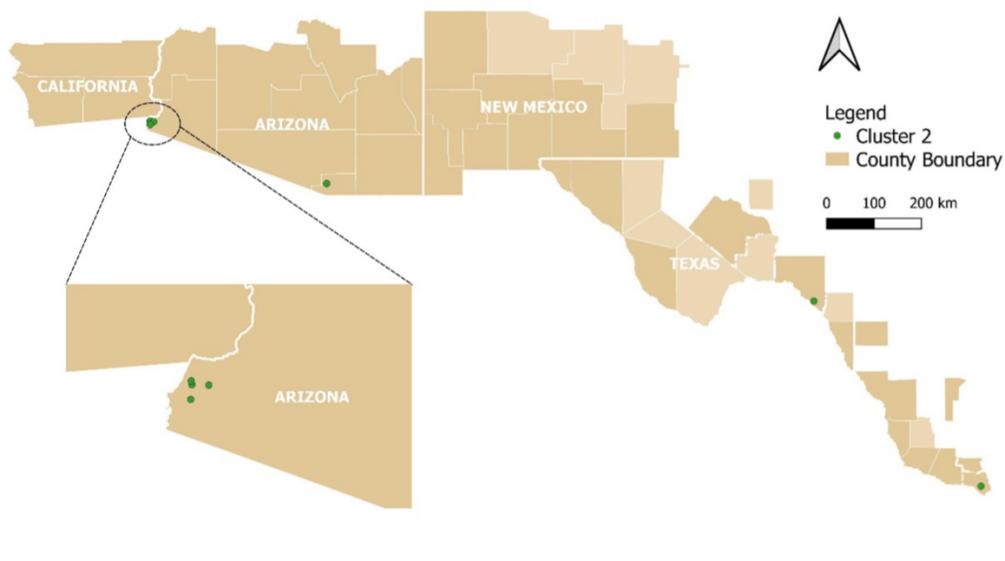

(c) Distribution of colonias in Cluster 2.

**Figure 7.** Spatial distribution of colonias in clusters without public water services. Counties shown in brown contain colonia communities, while counties shown in tan do not contain colonias.

## 6. Discussion

This paper introduces a new data-driven approach that combines both exploratory analysis—including clustering and predictive analysis—and classification to understand the water accessibility issues within underserved colonia communities. By grouping colonias based on a composed similarity measure, we were able to derive clusters of colonias with similar water insecurity conditions in terms of water access, wastewater treatment, water quality, and potential exposure to water hazards. The original dataset was hierarchically partitioned into multiple clusters, with two major subsets: colonias with public water supply services and those without. In general, we considered the former group to have fewer water-related issues than the latter. The results show that none of the clusters in the former group have reported water quality issues, and the majority of the communities have adequate water services in terms of public water supply and wastewater treatment services. The colonias in the second group are more concerning, as most of them have only limited access to a public water system. One cluster (as Figure 7[b] shows) still relies on water hauling to get drinking water, and another cluster (as Figure 7[c] shows), while small, has a severe water quality problem that requires immediate attention to improve the living conditions of the people there.

The results of this analysis provide evidence-based information to aid decision-making processes when prioritizing and allocating limited resources to address water issues in underserved communities. Different from existing priority indexing methods developed manually by experts or stakeholders, our approach provides an automated, interpretable, and reproducible workflow to understand common characteristics of colonia communities that are widely dispersed geographically. This solution is especially helpful in understanding complex water insecurity scenarios in which multiple factors must be considered simultaneously. This is also when it becomes very difficult for human experts to discern and develop comprehensive rules for determining resource allocation priorities.

Instead of serving as the de facto solution adopted directly by decision makers, the results presented in this work will become important supporting evidence for complex decision-making. For instance, we have compared our derived results with the RCAP rules (RCAP, 2015) that provide the priority rankings of colonias in terms of their infrastructure needs. The first and the second columns in Table 7 list the RCAP priority rankings (column 1, with "Priority 1" indicating colonias with the most urgent need for water infrastructure and access) and the corresponding rules defined by experts (column 2). Column 3 shows the machine-generated clusters from our research and how they map in relation to the RCAP priority rankings. It can be seen that multiple clusters (e.g., clusters 6, 2, and 10) from our results share the same priority level as the RCAP ranking (i.e., Priority 1). Because our analysis derives the distinctive challenges that colonias in different clusters are facing, these results could offer new insights beyond the expert-defined rules. For example, according to the mapping table (Table 5), clusters 6, 2, and 10 all belong to Priority 1 in terms of their infrastructure needs; among these, clusters 6 and 2 are facing health hazards related to water, and the residents of the colonias in both clusters do not have access to either public wastewater or public water services, instead relying on private wells as the main water source. In contrast, the residents of the colonias in Cluster 10, although small by population, must obtain water mainly by hauling. By the RCAP definition, however, all the colonias in clusters 6, 2, and 10 have a Priority 1 ranking, making it difficult to discern the characteristics and differences among them. As such, the varying characteristics outlined in the clusters can provide decision makers with a more detailed view of the water access and infrastructure needs of these colonias.

Our results also show that clusters 0, 7, 8, 12, and 13 all belong to Priority 2 according to the RCAP ranking. Among these colonias, clusters 0, 7, and 12 have no health hazard concerns, and although they do not have public water supply or wastewater treatment services, most have their own private wells. However, the colonias in clusters 8 and 13 have no clear water source description, meaning that their water access conditions need further investigation.

**Table 7.** Mapping of clustering results in relation to RCAP priority rankings according to the need for water infrastructure (bolded cluster numbers indicate those containing colonias with public water supply services).

| Priority | Requirements | Machine-generated clusters |
|---|---|---|
| Priority 1 | No wastewater service, and health hazard may be present | Cluster 2, Cluster 6, Cluster 10 |
| Priority 2 | No public water service and no health hazard, OR no wastewater service, service not adequate, and no health hazard, OR residents served by public water and wastewater systems but one or both are in serious violation of regulations | Cluster 0, Cluster 7, Cluster 8, Cluster 12, Cluster 13, Cluster **11**, Cluster **22** |
| Priority 3 | Not served by a publicly owned water AND/OR No access to wastewater service AND No wastewater service, but one is in the process of being financed | Cluster **5** |
| Priority 4 | Served by public water facilities and not served by public wastewater service, but wastewater disposal systems appear to be adequate or served by both public water service and publicly owned wastewater facilities | Cluster **1**, Cluster **21**, Cluster **19**, Cluster **9**, Cluster **4**, Cluster **20** |
| Priority 5 | No inhabitants | Cluster 3 |

Colonias with public water services are in a better condition than are those without when the water supply is safe. We used maximal matching (i.e., of where most of the colonias fall in the RCAP priority) to map them into the table. The cluster numbers of those containing colonias with public water services are shown in bold in Table 7. The majority of these clusters have a lower priority (with most being categorized by the RCAP as Priority 4), while only a couple of such clusters (i.e., 11 and 22) fall in RCAP Priority 2, as the majority of the colonias in these clusters do not have access to wastewater treatment systems and their existing services have been identified as inadequate. This detailed categorization of water infrastructure needs, made possible through our analysis, provides decision makers with a more comprehensive view of the water access insecurity challenges facing the colonias.

This work has significant policy implications. Water insecurity in the Global North is a phenomenon that has only recently been recognized (Meehan et al., 2020a). In the US, much is still unknown about the drivers of water insecurity at the community level and how they can be addressed (Meehan et al., 2020b). The methods presented here enable researchers to efficiently identify clusters of water insecurity conditions, which can facilitate analysis and intervention. For policymakers, knowing which colonias are dealing with unacceptable water access conditions can help them in supporting targeted policy solutions, including increasing funding and social and technical support. As experiences across colonias vary, this will be of particular importance for colonias that do not have access to public water systems, those that rely on water hauling, and those that face water quality challenges. Therefore, for congressional policymakers in need of data to support proposed investments in water infrastructure to improve local conditions (Deitz et al., 2019), this work provides strong evidence about which colonias should be prioritized. Additionally, this work can assist local policymakers across the four border states in developing policies to fund cross-sector collaborations with different entities (e.g., non-profit and for-profit water organizations) to increase the successful implementation of long-term solutions for water equity in these underserved communities.

## 7. Conclusion

Colonias communities are facing significant water access inequality. Despite the decades of data collection that has been performed in relation to the drinking and sanitation water infrastructure, water treatment, and water quality in the colonias, little effort has been exerted to quantitatively assess and rank the progress made to reduce these historical water inequity issues. Prior assessments have relied on interviews, stakeholder group discussions, and rule-based assignments to rank water issues across different colonias (Arsenault, 2021). This is a time-consuming process and is therefore unsuitable for addressing large-scale problems. This paper introduces a combined machine learning approach to identify colonias at different severity levels of water insecurity. It not only identifies colonias with similar water access conditions (Tables 5 & 6), but it also provides rankings on the importance of factors that characterize different groups of colonias (Figures 5 & 6). Although geolocation information is not used in the analysis, as we hope to gain an across-the-board understanding of the water insecurity challenges among all the colonias communities, our results (e.g., Figure 7) have clear geographic interpretations, which is critical for assessing water insecurity.

Our research fills the knowledge gap in the lack of understanding of the actual water access issues in the colonias; based on data-driven analysis, it is capable of identifying colonias groups across states facing different levels of water-related problems, which will help researchers and decision makers to (1) gain a big-picture view of the water problems, (2) collaboratively develop customized solutions to increase water access/quality, and (3) share and apply the solutions across state borders for colonias groups with similar needs. The proposed analytical framework can be easily scaled up to perform an automated data analysis to better understand water access problems nationwide. Another significant strength of this method is that the machine learning–based decision tree analysis helps to reveal how the machine generates the clustering results, which increases the model's interpretability and decision makers' confidence in machine-aided decisions. Our methodological workflow is also reproducible and can be easily adapted to different geospatial analysis and decision scenarios (e.g., to analyze data state by state vs. as one large dataset, or to analyze one dimension or different combinations of water insecurity issues). Hence, the proposed solution successfully tackles the common concern of the lack of explainability when using machine learning models in real-world applications (Goodchild & Li, 2021).

In the future, we will integrate multi-source data into this analysis and extend our research in a more detailed study of water insecurity issues at the neighborhood level. By collaborating with social scientists, civil engineers, legislators and policymakers, as well as local governments, we aim to develop sustainable solutions to create a combined social, physical, and cyber infrastructure that can address water insecurity issues not only in colonias but in all underserved communities.

## 8. Data and code availability

The data and code used for this research is openly accessible on GitHub: https://github.com/ASUcicilab/Colonias_Water_ML. README.md provides instructions on how to run the code for the analytical workflow.